\documentstyle[12pt,times,epsf,ifthen,overcite]{article}
\def\thefootnote{\fnsymbol{footnote}}
\newcommand{\spaper}[5]{
\ifthenelse{\equal{#1}{}}{}{#1, }#2 {\bf #3}, #5 (19#4)}
\newcommand{\db}{\hspace{-0.2ex}\not\hspace{-0.7ex}D\hspace{0.1ex}}
\newcommand{\comenta}[1]{}
\newcommand{\be}{\begin{equation}}
\newcommand{\ee}{\end{equation}}
\newcommand{\rms}{\rm\scriptsize}
\newcommand{\eqn}[1]{(\ref{#1})}

\newcommand{\bear}{\begin{eqnarray}}
\newcommand{\eear}{\end{eqnarray}}

\newcommand{\Tr}[1]{{\mathrm{Tr}}\{#1\}}
\begin{document}

\begin{titlepage}

\vspace*{-0.75in}
\rightline{\bf FTUV/96-85}
\rightline{\bf IFIC/96-94}
\vspace{2cm}

\begin{center}
{\huge \bf Hard $m_t$ Corrections as a Probe of the 
\\[0.35cm] Symmetry Breaking Sector}

\vspace{1.25cm}
{\large
{\bf J. Bernab\'eu, D. Comelli,
A. Pich} and {\bf A. Santamaria}\\[0.5cm]
Departament de F\'{\i}sica Te\`orica, IFIC,\\
CSIC-Universitat de Val\`encia\\
46100 Burjassot, Val\`encia, Spain\\
}

\vspace{2cm}
{\large\bf Abstract}
\end{center}
\begin{quotation}
Non-decoupling effects related to a large $m_t$ affecting
non-oblique radiative corrections in vertices ($Z\bar{b}b$)
and boxes ($B$--$\bar{B}$ mixing and $\epsilon_K$) are very sensitive
to the particular mechanism of spontaneous symmetry breaking. We analyze
these corrections in the framework of a chiral electroweak standard model
and find that there is only one operator in the effective lagrangian which
modifies the longitudinal part of the $W^+$ boson
without touching
the oblique corrections. The inclusion of this operator affects the
$Z\bar{b}b$ vertex, the $B$--$\bar{B}$ mixing and the CP-violating
parameter $\epsilon_K$, generating interesting correlations among the
hard $m_t^4 \log m_t^2$ corrections to these
observables, for example, the maximum vertex $Z\ b\bar{b}$ correction
allowed by low energy physics is about one percent.
\end{quotation}
\vspace{2cm}
\leftline{\today}

\end{titlepage}

\def\baselinestretch{1.1}

One of the basic ingredients of the standard model (SM)
is the spontaneous breaking of the 
electroweak gauge symmetry. In the SM it is implemented through the 
Higgs mechanism
in which the would-be Goldstone excitations are absorbed into the longitudinal
degrees of freedom of the gauge bosons.
The spontaneous symmetry breaking (SSB) is realized linearly, 
that means, by the use of
a scalar field which acquires a non-zero vacuum expectation value.
The spectrum of physical particles contains then
not only the massive vector bosons but also a neutral scalar Higgs field
which must be relatively light.

In a more general scenario, the SSB can be parametrized in terms of a
non-renormalizable lagrangian which contains the SM gauge symmetry
realized non-linearly\cite{nonlinear,longhitano}~. This non-linearly realized SM is
also called the chiral realization of the SM (\def\csm{$\chi$SM~}\csm) due to 
of its similarity with low energy QCD chiral lagrangians.
It includes, with a particular choice of the parameters
of the lagrangian, the SM, as long
as the energies involved are small compared with the Higgs mass which
is not present in the effective Lagrangian. In addition it can also 
accommodate any model that reduces to the SM at low energies as
happens in many technicolour scenarios.
The price to be payed for this general parametrization is the lose
of renormalizability and, therefore, the appearance of many couplings which
must be determined from experiment or computed in a more fundamental theory.

Since the SSB is related to the bosonic sector, one would expect that any
deviation from the SM SSB mechanism would affect especially the gauge boson 
propagation properties,
the so-called oblique corrections, which are parametrized in terms of the
S,T,U parameters\cite{stu} (or the $\epsilon_1,\epsilon_2$ and $\epsilon_3$
pameters\cite{altarelli}). In fact these corrections have been extensively
studied in the framework of the
\csm~\cite{feruglio}~. In particular, one would think that one should look into
quantities which are $M_H$-dependent in the SM to test the SSB sector.
However, it is interesting to realize that the only $M_H$-dependent radiative
correction, $\Delta\rho$, has an agreement with the SM prediction at the
per-mil level. Vertex corrections, whose $M_H$ dependence appears only
at the two loop level, are not so well known\footnote{And, in fact,
in the last years there has been a big controversy about the  
$R_b=\Gamma(Z\rightarrow b\bar{b})/\Gamma(Z\rightarrow hadrons)$ 
value.}.

On the other hand, the would-be Goldstone
bosons coming from SSB also couple to fermions. In fact, all non-decoupling
effects of the SM related to a large top quark mass, $m_t$, come 
from the coupling of the
would-be Goldstone bosons to the top quark. Therefore, we can
expect any non-decoupling quantity related to a heavy top-quark to be
sensitive to the would-be Goldstone boson propagation properties and
couplings, that is, to the specific mechanism of SSB. 

In the SM, large
$m_t^2$ effects appear, in addition to the oblique corrections, in the
vertex $Zb\bar{b}$, that is in $R_b=\Gamma_b/\Gamma_{h}$, and in
$B$--$\bar{B}$ and $K$--$\bar{K}$
mixing\footnote{Of course, non-decoupling effects appear in
other observables, but only in the quantities we just mentioned present 
experiments are sensitive enough to see the effects}. Then, 
we will use these quantities
to explore possible deviations of the SM spontaneous symmetry breaking
mechanism. To do so we will use a \csm only for the bosonic sector of the 
theory and leave fermion couplings as in the linear SM.

It turns out that there is only one operator in the effective lagrangian
that affects the $Zb\bar{b}$ vertex without touching the oblique corrections
(which as commented before agree with the SM at the per-mil level). This
operator modifies the propagation properties of the charged would-be
Goldstone bosons, that is, the longitudinal component of the $W^+$ boson.
Therefore, it will also affect any observable in which the non-decoupling
effects of a large $m_t$ are important, in particular
$B$--$\bar{B}$ mixing, and $\epsilon_K$.

In the non-linear realization of the SM 
the Goldstone bosons $\pi^a$ associated to the SSB of $SU(2)_L \times SU(2)_R
\rightarrow SU(2)_{L+R}$  are collected in
a matrix field $U(x)=\mathrm{Exp}\{i\, \pi^a\tau^a/v\}$.
The operators in the effective chiral
Lagrangian are classified according to the number
of covariant derivatives acting on $U(x)$.

The lowest-order operators just fix the values of the $Z$ and $W$ mass at
tree level and do not carry any information on the underlying physics.
Therefore, in order to extract some 
information on new physics we must start  studying the effects coming 
from higher-order operators.
Departure of those coefficients from the SM predictions 
can be a hint  for  the existence of new physics.

The lowest order effective chiral
lagrangian can be written in the following
way:
\be
L=L_B+L_{\psi}+L_Y~,
\ee
where
\be
L_B= -\frac{1}{2}
\Tr{\hat{W}_{\mu\nu}\hat{W}^{\mu\nu}+\hat{B}_{\mu\nu}\hat{B}^{\mu\nu}}
+\frac{v^2}{4}\Tr{D_{\mu}U^+D^{\mu}U}~,
\ee
with
$\hat{W}_{\mu\nu}=W^a_{\mu\nu}\tau^a/2\;\;\;\hat{B}^{\mu\nu}=B^{\mu\nu}\tau^3/2$,
and
$D^\mu U = \partial^\mu U + i\frac{g}{2} W_a^\mu \tau^a U-i\frac{g'}{2} B^\mu U \tau^3$.
$L_{\psi}$ is the usual fermionic kinetic lagrangian
and
\be
L_Y=-\bar{Q}_LUM_q Q_R+h.c~,
\ee
where $M_q$  is a $2 \times 2$ block-diagonal matrix containing the
$3 \times 3$ mass matrices of the up and down quarks and $Q_L$ and $Q_R$ are
doublets containing the up and down quarks for the three families
in the weak basis.

At the next order, that is containing at most four derivatives,
the  $CP$ and $SU(2)_L\otimes U(1)_Y$ invariant effective chiral Lagrangian 
with only gauge bosons and Goldstone fields,
is described by  the 15 operators reported in ref.~\citen{longhitano}:
$L=\sum_{i=0}^{14} a_i O_i$.

The usual oblique corrections are only sensitive to $a_0$,
$(a_8+a_{13})$ and $(a_1+a_{13})$.
On the other hand, the operators proportional to $a_2,\;a_3,a_9$ and $a_{14}$ 
parametrize  the effective non-abelian gauge couplings that are 
tested by LEP2.
All the other couplings remain not tested because they only
contribute to four-point Green functions ($a_4,a_5,a_6,a_7,a_{10}$)
or because, although quadratic in the Goldstone fields, 
they do not contribute to the one-loop oblique corrections
($a_{11},a_{12}$).
For instance, the operator proportional to $a_{11}$:  
\be
O_{11}= \Tr{\left(D_{\mu} V^{\mu} \right)^2}~,
\label{a11operator}
\ee
with
$ V_{\mu}= (D_{\mu} U) U^+$
and
$ D^{\mu}V_{\mu}= \partial^{\mu} V_{\mu} +i g[\hat{W}_{\mu},
V^{\mu}]$ 
generates corrections
to the two point Green function of the $W^+,\; Z$ and would-be
Goldstone bosons:
\bear
O_{11}&=& g^2 W_{\mu}^+  \partial^{\mu}\partial^{\nu}  W_{\nu}^-+
\frac{g_Z^2}{2} Z_{\mu} \partial^{\mu}\partial^{\nu} Z_{\nu}-
4 \pi^+ \frac{\partial^4}{v^2} \pi^- \nonumber \\
 &-&2 \pi_3 \frac{\partial^4}{v^2}\pi_3+
\frac{2 g}{v}W_{\mu}^+\partial^{\mu} \partial^2\pi^- +
\frac{2 g}{v} W_{\mu}^-\partial^{\mu}\partial^2\pi^+
+ \frac{2 g_Z}{v} Z_{\mu}^+\partial^{\mu}\partial^2 \pi_3
+ O(\phi^3)~.
\eear
However, all these 
interactions involve always the longitudinal 
components of the gauge bosons and so do not enter
directly into the $\epsilon_i$ parameters.
The same happens to the operators $O_{12}$ and $O_{13}$ which affect
only the longitudinal part of the neutral Z boson.

The effects of the operator $O_{11}$ can be seen more easily 
once we use the following equation of motion 
involving the operators of the Lagrangian to lowest order
\footnote{This is allowed in the effective Lagrangian, even at the one loop
level, as long as we keep only the dominant pieces. The use of the equations
of motion is equivalent to a redefinition of the fields which affects
only higher order operators in the effective Lagrangian}:
\be
D_{\mu} V^{\mu} = \frac{i}{v^2}D_{\mu} 
\left(\bar{Q}_L \gamma^{\mu} \tau^a Q_L \tau^a\right)~,
\ee
\be
i\db Q_L = U M_q Q_R\ , \hspace{1cm} 
i\db Q_R = M_q^+ U^+ Q_L~.
\ee
Then the operator $O_{11}$ can be rewritten as 
\be
O_{11}= \frac{g^4}{8 M_W^4}
\left[\bar{Q} (\tau^a U M_q P_R- M_q^+ U^+\tau^a P_L ) Q\right]^2~,
\label{o11weak}
\ee
where $P_L$ and $P_R$ are the left and right chirality projectors.
By writing \eqn{o11weak} in terms of the mass eigenstates and
keeping only the terms proportional to the top quark mass we obtain
\be
O_{11}= \frac{g^4}{8 M_W^4} m_t^2\left((\bar{t}\gamma_5t)^2 -4
\sum_{f,f'}^{d,s,b}(\bar{f'}_L t_R)(\bar{t}_Rf_L) V_{tf} V^*_{t f'}\right)~.
\label{o114f}
\ee
Therefore, the effect to lowest order of 
the modification of the would-be Goldstone propagator can be written as
a four-fermion interaction proportional to quark masses.
This kind of operators appears also in the analysis of new physics with an
effective Lagrangian with SSB realized linearly\cite{linear}~.

Four fermion interactions are much more convenient for explicit calculations
and also to understand the effects of the new operator. For instance, it
is clear that the four-fermion interaction can only contribute to the
gauge-boson self-energies at two loops and therefore do not contribute
to the $\epsilon_i$ parameters at one loop. 

We discuss now some observables affected by the new interaction.

\vspace*{0.4cm}
{\large \it \underline{$R_b$} .--}
\vspace{0.5cm} 

We start with the evaluation of the corrections to the $Z\bar{b}b$ vertex.
We parametrize the effective $Z\bar{b}b$ vertex as:
\be
\frac{g}{c_W}Z^{\mu}\left(g_L^b \bar{b}_L\gamma_{\mu}b_L+
g_R^b \bar{b}_R\gamma_{\mu}b_R\right)]~,
\ee
with the values of the tree level couplings, $g_L^b =-1/2+s_W^2/3$
and $g_R^b=s_W^2/3$.

At one loop we parametrize the effect of new physics as a shift in the 
couplings:
\be
g_{L,R}^b\rightarrow g_{L,R}^b+\delta g_{L,R}^b~.
\ee
We calculate the one-loop contribution of the operator $O_{11}$
keeping only the divergent logarithmic piece. This means we neglect any possible
local contribution from the chiral lagrangian at order $p^6$. The relevant
diagram\footnote{The same result is of course obtained using the original
form (\ref{a11operator}) for $O_{11}$, where the effect of this operator
appears as a modification of the longitudinal $W$ propagator. However,
one needs to consider a larger number of Feynman diagrams in this case.}
is depicted in fig.~(1.a) and the result is: 
\be
\delta g_L= -\frac{\alpha}{4 \pi s_w^2} a_{11}
\frac{g^2}{4}\frac{m_t^4}{M_W^4}
\log \frac{\Lambda^2}{m_t^2}~.
\label{gl-shift}
\ee
A shift in the $Z b\bar{b}$ couplings gives a shift in $R_b$ given
by
\be
R_b=R_b^{SM}\frac{1+\delta^{NP}_{bV}}
{1+R_b^{SM} \delta^{NP}_{bV}}~,
\label{rbtodelta}
\ee
with
\be
\delta^{NP}_{bV} = \frac{\delta \Gamma_b}{\Gamma_b^{SM}}
\approx 2\frac{g_L^{b}}
{(g_L^b)^2+(g_R^b)^2}\delta g_L^b =-4.58\, \delta g_L^b~. \label{rb-shift}
\ee

The ALEPH collaboration has presented
a new analysis of $R_b$ data which leads to results which are compatible with 
the
standard model predictions at the one sigma level\cite{polland1}~. In fact
the new world average is\cite{polland2} $R_b = 0.2178 \pm 0.0011$ 
to be compared with the SM expectation for
$m_t=175$~GeV $R_b^{SM} = 0.2157 \pm 0.0002$. 
Clearly the new
value of $R_b$ is within two standard deviations of the standard model
predictions\cite{sm}~. 

Using these data on $R_b$ we get 
\begin{equation}
\delta^{NP}_{bV} =0.012\pm 0.007~.
\label{shift-number}
\end{equation}

\vspace*{0.4cm}
{\large \it \underline{$K$--$\bar{K}$ and $B$--$\bar{B}$ Mixing} .--}
\vspace{0.5cm}

In the SM, the mixing between the $B^0$ meson and its antiparticle is 
completely dominated by the top contribution. The explicit $m_t$ dependence
of the corresponding box diagram is given by the loop function\cite{burasrev}
\be
S(x_t)_{\mbox{\rms SM}} = 
{x_t\over 4} \left[ 1 + {9\over 1 - x_t} - {6\over (1 - x_t)^2} 
- {6 x_t^2 \ln{x_t}\over (1 - x_t)^3}\right] ,
\qquad\quad x_t\equiv {\overline{m}_t^2\over M_W^2}~,
\ee
which contains the hard $m_t^2$ term, $S(x_t)\sim x_t/4$, induced
by the longitudinal $W$ exchanges.
The same function regulates the top--quark contribution to the 
$K$--$\bar K$ mixing parameter $\varepsilon_K$.
The measured top--mass, $m_t = 175\pm 6$ GeV 
[$\overline{m}_t \equiv\overline{m}_t(m_t) = 167 \pm 6$ GeV], implies
$S(x_t)_{\mbox{\rms SM}} = 2.40 \pm 0.13$.

The correction induced by the new operator, $O_{11}$,
can be parametrized as a shift on the function $S(x_t)$. The calculation
of the diagrams$^\thefootnote$ in fig.~(1.b) leads to the result: 
\be
S(x_t) = S(x_t)_{\mbox{\rms SM}}+\delta S(x_t)\, , \qquad\quad
\delta S(x_t) = -a_{11}\, \frac{g^2 m_t^4}{2 M_W^4}\, \ln{\Lambda^2\over m_t^2}
\ \ .
\ee
Thus, the hard $m_t^4 \ln{m_t^2}$ contributions to $\delta^{NP}_{bV}$ and
$\delta S(x_t)$ are correlated:
\be\label{eq:correlation}
\delta S(x_t) \, =\, {32 \pi^2\over |V_{tb}|^2 g^2} \delta g_L^b
\, = \, -163 \,\delta^{NP}_{bV} .
\ee

We can use the measured $B^0_d$--$\bar B^0_d$ mixing\cite{gibbons}~,
$\Delta M_{B^0_d} = (0.464 \pm 0.018) \times 10^{12}\,\mbox{\rm s}^{-1}$, 
to infer the experimental value of $S(x_t)$ and, therefore, to
set a limit on the $\delta g_L^b$ contribution.
The explicit dependence on the quark--mixing parameters can be resolved
putting together the constraints from $\Delta M_{B^0_d}$, $\varepsilon_K$
and $\Gamma(b\to u)/\Gamma(b\to c)$.
Using the  Wolfenstein parametrization\cite{WO:83}
of the quark--mixing matrix, one has:
\be\label{eq:mixing}
\left|\frac{V_{td}}{\lambda V_{cb}}\right| \, = \,
\sqrt{(1-\rho)^2+\eta^2} \, = \,
{(1.21\pm 0.09)\over \sqrt{S(x_t)}}\, {185\:\mbox{\rm MeV}\over
\sqrt{\eta_B}\, (\sqrt{2} f_B \sqrt{B_B})} \, = \,
{(1.21^{+0.50}_{-0.30}) \over\sqrt{S(x_t)}} \, ,
\ee
\be\label{eq:epsilon}
\eta\left[(1-\rho)\, A^2 \eta_2\, S(x_t)+P_0\right]A^2 B_K\, =\, 0.226
\ee
\be\label{eq:btou}
\left|\frac{V_{ub}}{\lambda V_{cb}}\right| \, = \,
\sqrt{\rho^2+\eta^2} \, = \, 0.36 \pm 0.09 \, .
\ee
We have taken
$\lambda \equiv |V_{us}| = 0.2205\pm 0.0018$,
$|V_{cb}|\equiv A \lambda^2 = 0.040\pm 0.003$
and
$|V_{ub}|/|V_{cb}| = 0.08 \pm 0.02$.
The numerical factor on the rhs of eq.~\eqn{eq:mixing} should be
understood as an allowed range, because the error is dominated by the
large theoretical uncertainties in the hadronic matrix element of the
$\Delta B=2$ operator; it corresponds to\cite{buras,prades}
$\sqrt{\eta_B}\, (\sqrt{2} f_B \sqrt{B_B}) = (185\pm 45)$ MeV.
In eq.~\eqn{eq:epsilon}, $\eta_2 = 0.57\pm 0.01$ is the short--distance QCD
correction\cite{BJW:90}~, 
while $P_0=0.31\pm 0.02$ takes into account the charm 
contributions\cite{buras}~. 
For the $\Delta S=2$ hadronic matrix element we have chosen
the range\cite{prades}
$B_K = 0.6\pm 0.2$.

Both the circle \eqn{eq:mixing} and the hyperbola \eqn{eq:epsilon}
depend on the the value of $S(x_t)$.
The intersection of the two circles \eqn{eq:mixing} and \eqn{eq:btou}
restricts $S(x_t)$ to be in the range
$0.39 < |S(x_t)| < 9.7$.
The request of simultaneous intersection with the hyperbola 
$\epsilon_K$ imposes a further constraint.
Since a positive value of $B_K$ is obtained by all present calculations
and $S(x_t)_{\mbox{\rms SM}}>0$, the SM 
implies a positive value for $\eta$.
In our case, the constraint that the total  
$S(x_t)=S(x_t)_{\mbox{\rms SM}}+\delta S(x_t)$
is  positive does not exist and this opens the possibility of solutions
also with $\eta<0$;
however, this would imply a huge correction $\delta S(x_t)$.
Taking $\eta>0$, the three curves (bands) intersect if
$S(x_t) > S(x_t)_{\mbox{\rms min}} = 1.0$.

The minimum value of $S(x_t)$ is reached for
$V_{cd}^{\mbox{\rms max}}$, $B_K^{\mbox{\rms max}}$
and $|V_{ub}/V_{cb}|^{\mbox{\rms max}}$.
Taking a more conservative $\pm 0.14$ error in eq.~\eqn{eq:btou}
(corresponding to $|V_{ub}/V_{cb}| = 0.08\pm 0.03$) would result
in $S(x_t)_{\mbox{min}} = 0.8$.

The shift in $g_L^b$ required by $R_b$ [eq.~\eqn{shift-number}] and
the relation \eqn{eq:correlation} imply
\begin{equation}
\delta S=-2.0 \pm 1.1~,
\label{S-shift}
\end{equation}
i.e., $-0.7 < S(x_t) < 1.5$.
Thus, the present experimental measurements of $R_b$  and
the low-energy constraints from the
usual {\it unitarity triangle fits}
are compatible with the introduction of the operator $O_{11}$. 
From eq.~(\ref{eq:correlation})  and eq.~(\ref{shift-number}) and the constraint
$S\geq S_{min}=1$ we can see that the maximum (positive) value of $\delta_{bV}^{NP}$ 
allowed by low-energy physics is
\[
\delta_{bV}^{NP} < 0.01~, 
\]
which is even stronger than the values obtained by present direct measurements
of $R_b$ (eq.~(\ref{shift-number}).

\vspace{1.5cm}

We acknowledge interesting discussions with Misha Bilenky, Domenec Espriu and 
Joaquim Matias. One of us  (D.C.) is indebted to the Spanish 
Ministry of Education and Science for a postdoctoral fellowship. This work
has been supported by the Grant AEN-96/1718 of CICYT, Spain.

%

%
%%%%%%%%%%%%%%%
\def\mafigura#1#2#3#4{
  \begin{figure}[hbtp]
    \begin{center}
      \epsfxsize=#1 \leavevmode \epsffile{#2}
    \end{center}
    \caption{#3}
    \label{#4}
  \end{figure} }
%%%%%%%%%%%%%%%
\mafigura{12cm}{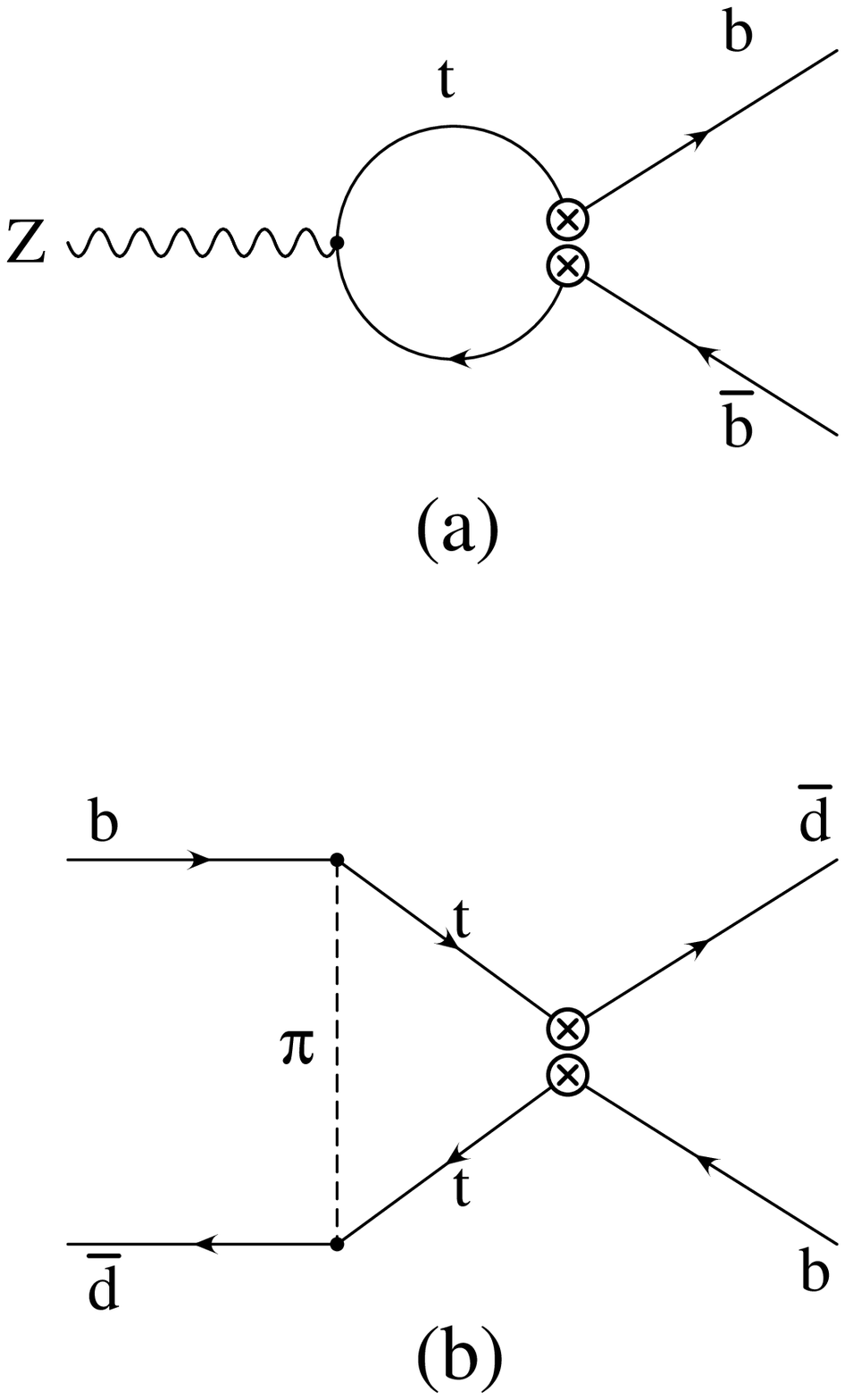}{a) Contribution of the effective operator $O_{11}$
to $Z\rightarrow b\bar{b}$. b) Contribution of the effective operator $O_{11}$
to $B$--$\bar{B}$ mixing}{fig1}
%%%%%%%%%%%%%%%

\end{document}